\documentclass[aps,prl,twocolumn,showpacs,superscriptaddress,groupedaddress]{revtex4-1}

\usepackage{graphicx}
\usepackage{dcolumn}
\usepackage{bm}
\usepackage{amsmath,amssymb}
\usepackage{float}

\usepackage{pgfplots}
\pgfplotsset{compat=1.18}
\usepackage{xcolor}
\usepackage{subcaption}

\definecolor{prlblue}{RGB}{33,113,181}
\definecolor{prlorange}{RGB}{217,95,2}
\definecolor{prlgreen}{RGB}{27,158,119}
\definecolor{prlgray}{gray}{0.45}

\pgfplotsset{
  prl/.style={
    width=\columnwidth,
    height=0.72\columnwidth,
    axis lines=left,
    line width=1.0pt,
    tick style={semithick},
    axis line style={semithick},
    xlabel style={font=\small},
    ylabel style={font=\small},
    tick label style={font=\footnotesize},
    legend style={font=\footnotesize, draw=none, fill=none, row sep=1pt},
    legend cell align={left},
    every axis plot/.append style={line width=1.2pt},
    cycle list={{
      prlblue, mark=*, mark size=2.2pt
    },{
      prlorange, dashed, mark=square*, mark size=2.2pt
    },{
      prlgreen, dashdotted, mark=triangle*, mark size=2.2pt
    }},
  }
}

\begin{document}

\preprint{APS/123-QED}

\title{Amorphous Solid Model of Vectorial Hopfield Neural Networks}

\author{Filippo Gallavotti}
\email{filippo.gallavotti@studenti.unimi.it}
\affiliation{Department of Physics ``A. Pontremoli", University of Milan, via Celoria 16, 20133 Milan, Italy}

\author{Alessio Zaccone}
\email{alessio.zaccone@unimi.it}
\affiliation{Department of Physics ``A. Pontremoli", University of Milan, via Celoria 16, 20133 Milan, Italy}

\begin{abstract}
We introduce a three-dimensional vector Hopfield model where each neuron is a unit vector on $S^2$ and synaptic weights are $3\times3$ blocks built by a vectorial Hebbian rule, analogous to the Hessian of amorphous solids. The resulting rigid energy landscape strongly stabilizes stored patterns. Simulations and spectral analysis show that the critical storage ratio grows roughly linearly with connectivity and enters a high-connectivity regime with enhanced capacity, a persistent spectral gap, and enlarged basins of attraction, yielding robust associative memory.
\end{abstract}

\maketitle

The Hopfield model \cite{Hopfield} is a cornerstone of associative memory
in neural networks and artificial intelligence \cite{jcrn-3nrc}. In its classical
formulation, $N$ binary neurons $s_i \in \{-1,+1\}$ evolve according to
symmetric interactions $W_{ij}$ defined by the Hebbian rule from stored
patterns. The dynamics minimizes
$E = -\frac{1}{2}\sum_{ij} W_{ij} s_i s_j$, leading to pattern retrieval
via convergence to energy minima. Connections between neural network models and disordered materials have
been explored in spin glasses \cite{Sompolinsky,Jesi,Nicoletti}, jammed
packings \cite{Wyart,Martiniani}, random resistor and memristive
networks \cite{resistors}, and elastic networks \cite{Altman}. Recent
work has emphasized the relation between the ``cost landscape'' and the
physical landscape encoded by the cost Hessian and physical Hessian,
respectively \cite{Stern_PRL}.

Neural networks and amorphous solids share structural similarities in
their interaction matrices and in the emergence of multiple metastable
states. In amorphous solids the Hessian describing elastic interactions
exhibits a random block structure analogous to weight matrices in
neural networks with vectorial degrees of freedom.

Motivated by this analogy, we introduce a multi-dimensional extension of
the Hopfield model. Each neuron is a unit vector
$\mathbf{s}_i \in \mathbb{R}^3$, and interactions are built through outer
products following a vectorial Hebbian rule. This uses random block
matrices previously introduced for amorphous solids
\cite{zaccone2011approximate,Cicuta} and only recently adopted in neural
network theory \cite{Nicoletti}.

The vectorial extension is relevant to systems with orientational degrees of freedom, such as liquid crystals, polymeric glasses, biological networks such as the cytoskeleton, and enables systematic comparison with classical binary Hopfield networks \cite{hertz}. We
start by reviewing the binary Hopfield model.

\begin{figure}[tb]
\includegraphics[width=\columnwidth]{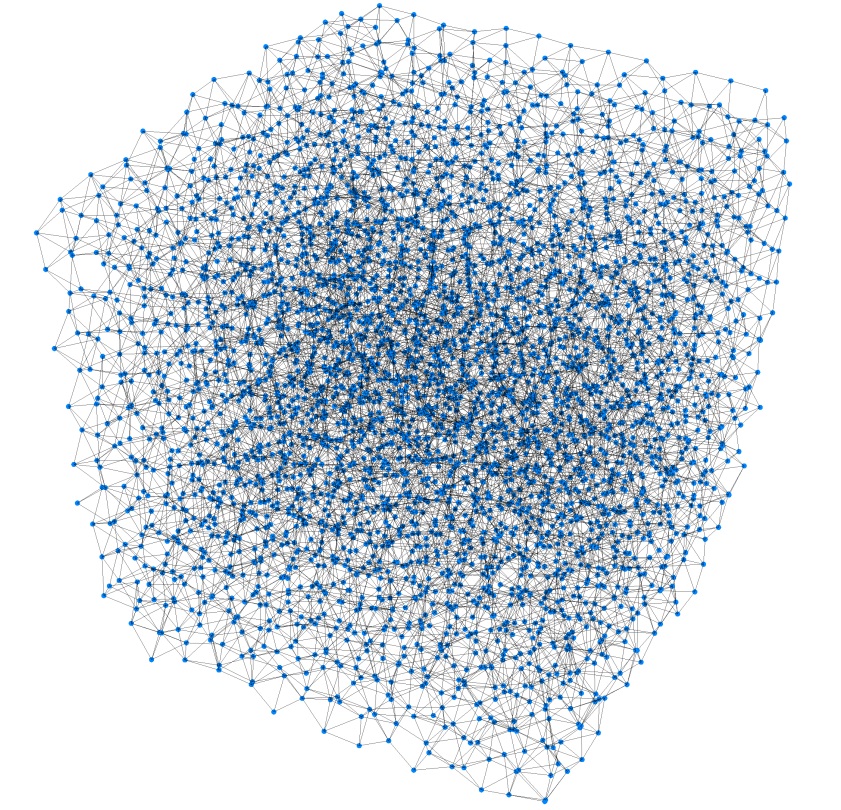}
\caption{Rendering of a 3D network of $N$ nodes with average
coordination number $Z=7$ and nearly uniform orientation distribution.}
\label{fig1}
\end{figure}

We consider 3D networks of $N$ nodes (neurons) connected by 3D vectors
(``bonds'' or synapses) as in Fig.~\ref{fig1}. In the amorphous solid
analogy, the connections are springs with spring constant $\kappa$
\cite{Milkus2016,Milkus2017,Zaccone_book}.

The binary Hopfield model consists of $N$ neurons with
$s_i \in \{-1,+1\}$. To store $P$ patterns
$\{\boldsymbol{\xi}^{\mu}\}_{\mu=1}^P$ with
$\boldsymbol{\xi}^{\mu} = (\xi_1^{\mu}, \ldots, \xi_N^{\mu})$ and
$\xi_i^{\mu} \in \{-1,+1\}$, the weight matrix is
\begin{equation}
W_{ij} = \frac{1}{N} \sum_{\mu=1}^P \xi_i^{\mu} \xi_j^{\mu}\label{weight}
\end{equation}
with $W_{ii} = 0$ and $W_{ij} = W_{ji}$. The dynamics follows
asynchronous updates
$s_i \leftarrow \text{sign}\bigl(\sum_j W_{ij} s_j\bigr)$, which
minimize $E = -\frac{1}{2}\sum_{ij} W_{ij} s_i s_j$. Neuron $i$ updates
according to its total input, with $\Delta E \le 0$, and pattern
retrieval occurs when the system converges from partial or noisy initial
states.

In the vectorial model, each neuron is a unit vector
$\mathbf{s}_i \in \mathbb{R}^3$ with $|\mathbf{s}_i| = 1$. We store $P$
patterns $\{\boldsymbol{\xi}^{\mu}\}_{\mu=1}^P$, where each pattern is a
set of $N$ random unit vectors on $S^2$:
\begin{equation}
\boldsymbol{\xi}^{\mu} = (\boldsymbol{\xi}_1^{\mu}, \ldots, \boldsymbol{\xi}_N^{\mu}), \quad |\boldsymbol{\xi}_i^{\mu}| = 1.
\end{equation}
The weight matrix is defined via a vectorial Hebbian rule using outer
products:
\begin{equation}
\mathbf{W}_{ij} = \frac{1}{N} \sum_{\mu=1}^P \boldsymbol{\xi}_i^{\mu} \otimes \boldsymbol{\xi}_j^{\mu}
\label{eq:vectorial_hebbian}
\end{equation}
where $\otimes$ denotes the outer product, yielding $3 \times 3$ blocks
$\mathbf{W}_{ij}$. The full matrix $\mathbf{W}$ is $3N \times 3N$ with
block structure \cite{Cicuta}.

This parallels the Hessian in amorphous solids \cite{zaccone2011approximate}. For elastic interactions between particles at positions $\mathbf{r}_i$,
\begin{equation}
H_{ij}^{\alpha\beta} = \kappa_{ij} n_{ij}^{\alpha} n_{ij}^{\beta}
\end{equation}
with spring constant $\kappa_{ij}$ and bond unit vector
\begin{equation}
    \mathbf{n}_{ij}=(\sin \theta \cos \phi, \sin \theta \sin \phi, \cos \theta),\label{unit}
\end{equation}
where $\alpha,\beta =\{x,y,z\}$. Self-interactions are ignored. The
structure matches Eq.~(\ref{eq:vectorial_hebbian}), with stored
patterns playing the role of bond directions.

The energy is
\begin{equation}
E(\mathbf{s}) = -\frac{1}{2} \sum_{i,j=1}^N \mathbf{s}_i^T \mathbf{W}_{ij} \mathbf{s}_j \label{energy}
\end{equation}
and the dynamics preserves $|\mathbf{s}_i|=1$ by normalizing the local
field $\mathbf{h}_i = \sum_{j \neq i} \mathbf{W}_{ij} \mathbf{s}_j$ via
$\mathbf{s}_i^{\text{new}} = \mathbf{h}_i / |\mathbf{h}_i|$.

Simulations show reliable retrieval: from random initial configurations, dynamics converges to stored patterns (or complements) within $\sim 30$ asynchronous updates.

To construct the synaptic matrix, we use a pseudo-inverse learning rule
instead of standard Hebbian learning. The classical rule is optimal
only for mutually orthogonal patterns, which is not generic at finite
$P/N$. To enforce exact retrieval, we adopt a pseudo-inverse
generalization. Defining the pattern matrix
$\Xi = [\boldsymbol{\xi}^{(1)}\, \boldsymbol{\xi}^{(2)} \cdots \boldsymbol{\xi}^{(P)}]$
of size $3N \times P$, the Gram matrix of overlaps is
\begin{equation}   
G_{\mu\nu} = \frac{1}{N}\sum_{i=1}^{N} \boldsymbol{\xi}_i^{\mu} \cdot \boldsymbol{\xi}_i^{\nu},
\end{equation}
and its inverse $G^{-1}$ defines the pseudo-inverse operator
\begin{equation}
W = \frac{1}{N}\sum_{\mu,\nu=1}^{P} \Xi_{\mu}\, (G^{-1})_{\mu\nu}\, \Xi_{\nu}^{\top},
\end{equation}
which satisfies $W \Xi = \Xi$, so each stored pattern is an exact
eigenvector of $W$ with eigenvalue $1$. This guarantees perfect recall
of all training patterns in the fully connected case and provides a
natural extension of the classical formulation to correlated patterns.

We next study how memory capacity depends on network connectivity. We
introduce a $Z$-connected topology, where each neuron is symmetrically
coupled to its $Z/2$ nearest neighbours on either side along a circular
ring. The coordination number $Z$ interpolates between fully connected
($Z=N-1$) and sparse ($Z \ll N$) networks. The adjacency matrix
$C_{ij}$ satisfies
\[
C_{ij} =
\begin{cases}
1, & \text{if } j \in \text{neigh}(i) \text{ or } i \in \text{neigh}(j), \\
0, & \text{otherwise}.
\end{cases}
\]
Synaptic weights are built locally by solving, for each neuron $i$, a
regularized linear system enforcing consistency between target outputs
$\boldsymbol{\xi}_i^{\mu}$ and inputs from neighbours
$\boldsymbol{\xi}_j^{\mu}$ in all stored patterns. With
$\mathcal{N}_i$ the neighbour set of $i$, the local matrices
$\{W_{ij}\}_{j \in \mathcal{N}_i}$ satisfy
\[
\sum_{j \in \mathcal{N}_i} W_{ij} \boldsymbol{\xi}_j^{\mu} \simeq \boldsymbol{\xi}_i^{\mu}, \qquad \forall \mu = 1,\dots,P,
\]
via the ridge-regularized least-squares solution
\[
W_i = (X_i X_i^{\top} + \lambda I)^{-1} X_i Y_i^{\top},
\]
where $X_i$ collects neighbour states and $Y_i$ the target states for
neuron $i$. The parameter $\lambda$ ensures numerical stability and
controls smoothness. This local pseudo-inverse generalizes the
fully connected rule to sparse graphs while encoding topology.

\begin{figure}[htb]
\centering
\begin{subfigure}{0.49\textwidth}
\centering
\begin{tikzpicture}
\begin{axis}[
  width=\linewidth,
  height=0.78\linewidth,
  xmin=0, xmax=26,
  ymin=-0.005, ymax=0.135,
  grid=both,
  xlabel={$Z$ (connection number)},
  ylabel={$\gamma_c$ (critical storage ratio)},
  ticklabel style={/pgf/number format/fixed},
  legend style={draw=none, fill=none, font=\small, at={(0.02,0.98)}, anchor=north west},
]
\addplot+[
  only marks, mark=*, mark size=2.5pt, blue,
  error bars/.cd, y dir=both, y explicit,
  error bar style={line width=0.8pt}, error mark=|,
] table[x=Z, y=gc, y error=sd] {
Z    gc         sd
2    0.0104167  0.0015
4    0.015625   0.0018
6    0.034375   0.0025
8    0.0416667  0.0022
10   0.05625    0.0021
12   0.0625     0.0024
14   0.071875   0.0021
16   0.0833333  0.0026
18   0.09375    0.0023
20   0.105208   0.0021
22   0.119792   0.0025
24   0.125      0.0022
};
\addlegendentry{simulation mean $\pm$ sd}
\addplot[red, thick, domain=2:24, samples=2] {0.0053*(x - 0.166)};
\addlegendentry{$\gamma_c(Z)=0.0053\,(Z-0.166)$}
\node[anchor=north east, text=red, font=\small] at (rel axis cs:0.98,0.15)
{$R^2=0.996$};
\end{axis}
\end{tikzpicture}
\label{fig:gamma_vs_Z}
\end{subfigure}
\hfill
\begin{subfigure}{0.49\textwidth}
\centering
\begin{tikzpicture}
\begin{axis}[
  width=\linewidth,
  height=0.78\linewidth,
  xmin=0, xmax=65,
  ymin=-0.01, ymax=0.30,
  grid=both,
  xlabel={$Z$ (connection number)},
  ylabel={$\gamma_c$ (critical storage ratio)},
  ticklabel style={/pgf/number format/fixed},
  legend style={draw=none, fill=none, font=\small, at={(0.02,0.98)}, anchor=north west},
]
\addplot+[
  only marks, mark=*, mark size=2.5pt, blue,
  error bars/.cd, y dir=both, y explicit,
  error bar style={line width=0.8pt}, error mark=|,
] table[x=Z, y=gc_spec, y error=sd_spec] {
Z    gc_spec      sd_spec
2    0.0078125    0.0012
4    0.0182292    0.0015
6    0.0338542    0.0021
8    0.0390625    0.0018
10   0.0494792    0.0023
12   0.0546875    0.0020
14   0.0598958    0.0024
16   0.0703125    0.0021
18   0.0755208    0.0026
20   0.0755208    0.0022
22   0.0963542    0.0025
24   0.1015625    0.0023
26   0.1119792    0.0027
28   0.1276042    0.0024
30   0.1276042    0.0028
32   0.1432292    0.0025
34   0.1588542    0.0029
36   0.1640625    0.0026
38   0.1692708    0.0030
40   0.1796875    0.0027
42   0.1901042    0.0031
44   0.2005208    0.0028
46   0.2057292    0.0032
48   0.2109375    0.0029
50   0.2265625    0.0033
52   0.2421875    0.0030
54   0.2421875    0.0034
56   0.2630208    0.0031
58   0.2734375    0.0035
60   0.2786458    0.0032
62   0.2890625    0.0036
};
\addlegendentry{spectral estimate $\pm$ sd}
\addplot[red, thick, domain=2:62, samples=2] {0.0046*(x - 0.793)};
\addlegendentry{$\gamma_c(Z)=0.0046\,(Z-0.793)$}
\node[anchor=north east, text=red, font=\small] at (rel axis cs:0.98,0.15)
{$R^2=0.992$};
\end{axis}
\end{tikzpicture}
\label{fig:gamma_vs_Z_spectral}
\end{subfigure}

\caption{Critical storage capacity versus connectivity. Top figure: Dynamical
estimate of $\gamma_c$ with standard-deviation error bars across $5$
replicas. Bottom figure: Spectral estimate from the vanishing smallest eigenvalue
of the Hessian at each attractor. Both panels show an approximately
linear dependence of $\gamma_c$ on $Z$ with closely matching slopes.}
\label{fig:gamma_two_panels}
\end{figure}

Figure~\ref{fig:gamma_two_panels} shows that, for moderate connectivity,
the critical storage capacity $\gamma_c$ increases approximately
linearly with $Z$. A linear regression (for $N=64$ neurons, $T=500$
trials, and $L=1000$ maximum asynchronous updates) yields
\[
\gamma_c(Z) = 0.0053\,(Z - 0.17), \qquad R^2 = 0.996.
\]
Each additional connection reinforces the effective local field, thereby
enlarging basins of attraction: increasing $Z$ stiffens the associative
energy landscape, deepening and separating minima corresponding to
memorized states. The nearly linear growth reflects a mean-field balance
between stabilizing recurrent couplings and destabilizing cross-talk.

To corroborate this picture, we performed a spectral analysis of the
Hessian at each attractor, identifying $\gamma_c$ as the point where the
smallest eigenvalue $\lambda_{\min}$ vanishes, i.e.\ the loss of local
stability. The resulting relation
\[
\gamma_c(Z) = 0.0046\,(Z - 0.79), \qquad R^2 = 0.992,
\]
is in excellent agreement with the dynamical estimate. The consistency
between dynamical and spectral approaches shows that the transition from
stable to unstable retrieval corresponds to a spectral softening of the
Hessian, analogous to the vanishing lowest vibrational mode at jamming.
Thus the linear dependence of $\gamma_c$ on $Z$ supports an
interpretation of associative-memory stability as an emergent rigidity
phenomenon in a high-dimensional landscape.

\begin{figure}[H]
    \centering
    \label{Z_alti}
\begin{tikzpicture}
\begin{axis}[
  width=\linewidth,
  height=0.78\linewidth,
  xmin=0, xmax=65,
  ymin=-0.01, ymax=0.50,
  grid=both,
  xlabel={$Z$ (connection number)},
  ylabel={$\gamma_c$ (critical storage ratio)},
  ticklabel style={/pgf/number format/fixed},
  legend style={draw=none, fill=none, font=\small, at={(0.02,0.98)}, anchor=north west},
]
\addplot+[
  only marks, mark=*, mark size=2.5pt, blue,
  error bars/.cd, y dir=both, y explicit,
  error bar style={line width=0.8pt}, error mark=|,
] table[x=Z, y=gamma_c, y error=devstd] {
Z   gamma_c   devstd
2   0.0104167   0
4   0.0208333   0
6   0.0260417   1.5522e-10
8   0.0416667   0
10  0.0520833   3.10441e-10
12  0.0625      0.00360417
14  0.0677083   0.00294103
16  0.0833333   0.00245523
18  0.0885417   0
20  0.0989583   0.00245523
22  0.111979    0.00260417
24  0.125868    0.00194103
26  0.145833    1.24176e-09
28  0.15191     0.00194103
30  0.164062    0.00260417
32  0.171875    0
34  0.188368    0.00194103
36  0.203993    0.00194103
38  0.231771    0.00260417
40  0.248264    0.00245523
42  0.270833    0
44  0.296007    0.00194103
46  0.315972    0.00245523
48  0.336806    0.00245523
50  0.368056    0.00245523
52  0.396701    0.00467462
54  0.415799    0.00467462
56  0.431424    0.00194103
58  0.436632    0.00467462
60  0.445312    0.00397793
62  0.453125    0.00520833
};
\addlegendentry{simulation mean $\pm$ sd}
\addplot+[
  red, thick, no marks, domain=0:65,
] {0.0053*(x - 0.166)};
\addlegendentry{$\gamma_c(Z)=0.0053(Z-0.166)$}
\end{axis}
\end{tikzpicture}
\caption{High-connectivity regime of critical storage capacity.
Symbols: spectral estimate of $\gamma_c$ (mean $\pm$ sd) versus $Z$.
Red line: linear trend $\gamma_c(Z)=0.0053\,(Z-0.166)$ obtained at
moderate $Z$ (cf.\ Fig.~\ref{fig:gamma_two_panels}). The systematic
upward deviation at large $Z$ signals a crossover to an enhanced-capacity
regime where $\gamma_c$ exceeds the extrapolated low-$Z$ prediction.}
\end{figure}

Beyond the moderate-connectivity regime of Fig.~\ref{fig:gamma_two_panels},
Fig.~3 shows a crossover at large $Z$ where $\gamma_c$ rises
faster than the low-$Z$ linear trend $\gamma_c(Z)=0.0053\,(Z-0.166)$.
The deviation is systematic and becomes pronounced for $Z\gtrsim 30$,
indicating an enhanced-capacity regime. Increased local redundancy and
orientational constraints in the block-structured operator strengthen
effective fields and reduce cross-talk, effectively ``rigidifying'' the
associative landscape. This is consistent with the spectral picture,
where pattern modes remain separated from the bulk, indicating stronger
attractor stability as connectivity increases. While finite-size effects
cannot be completely excluded, the monotonic upward curvature with $Z$
points to a genuine connectivity-driven gain beyond the linear
mean-field regime.

We next examine how pattern orientation distributions affect memory
capacity by generating patterns with random orientations for
$\mathbf{n}_{ij}$. Because $\mathbf{n}_{ij}$ is parametrized by
$\{\theta,\phi\}$ [Eq.~\eqref{unit}], both $\theta$ and $\phi$ are drawn
from Gaussian distributions normalized over $[0,\pi]$ and $[0,2\pi]$,
respectively. In the limit $\sigma \rightarrow \infty$, the orientation
distribution in solid angle is uniform, with probability $1/4\pi$
\cite{zaccone2011approximate,ZBT}.

\begin{figure}[htb]
\centering
\begin{tikzpicture}
\begin{axis}[
  width=0.95\linewidth,
  height=0.7\linewidth,
  xmin=0, xmax=0.45,
  ymin=0.008, ymax=0.065,
  grid=both,
  major grid style={line width=0.3pt, draw=gray!50},
  minor grid style={line width=0.1pt, draw=gray!20},
  xmajorgrids=true,
  ymajorgrids=true,
  xminorgrids=true,
  yminorgrids=true,
  xlabel={Angular standard deviation $\sigma$},
  ylabel={Critical storage ratio $\gamma_c$},
  label style={font=\small},
  ticklabel style={/pgf/number format/fixed, font=\small},
  legend style={
    draw=none,
    fill=white,
    fill opacity=0.8,
    cells={anchor=west},
    font=\small,
    at={(0.02,0.98)},
    anchor=north west
  },
  legend image post style={mark=*, mark size=1.8pt},
  enlargelimits=true,
]
\addplot+[
  only marks, mark=*, mark size=2pt, blue, solid,
  error bars/.cd, y dir=both, y explicit,
  error bar style={line width=0.6pt, blue},
  error mark options={blue, line width=0.6pt},
] table[x=sigma, y=gc_Z4, y error=sd_Z4] {
sigma gc_Z4       sd_Z4
0.05  0.0104167   0
0.10  0.0104167   0
0.15  0.0104167   0
0.20  0.0104167   0
0.25  0.0125000   0.00255155
0.30  0.0114583   0.00208333
0.35  0.0156250   0
0.40  0.0156250   0
};
\addplot+[
  only marks, mark=*, mark size=2pt, red, solid,
  error bars/.cd, y dir=both, y explicit,
  error bar style={line width=0.6pt, red},
  error mark options={red, line width=0.6pt},
] table[x=sigma, y=gc_Z6, y error=sd_Z6] {
sigma gc_Z6       sd_Z6
0.05  0.0114583   0.00208333
0.10  0.0135417   0.00255155
0.15  0.0125000   0.00255155
0.20  0.0156250   0
0.25  0.0156250   0.00329404
0.30  0.0187500   0.00255155
0.35  0.0187500   0.00255155
0.40  0.0187500   0.00255155
};
\addplot+[
  only marks, mark=*, mark size=2pt, green!70!black, solid,
  error bars/.cd, y dir=both, y explicit,
  error bar style={line width=0.6pt, green!70!black},
  error mark options={green!70!black, line width=0.6pt},
] table[x=sigma, y=gc_Z8, y error=sd_Z8] {
sigma gc_Z8       sd_Z8
0.05  0.0145833   0.00208333
0.10  0.0166667   0.00208333
0.15  0.0187500   0.00255155
0.20  0.0177083   0.00255155
0.25  0.0229167   0.00255155
0.30  0.0218750   0.00208333
0.35  0.0260417   0.00329404
0.40  0.0270833   0.00389756
};
\addplot+[
  only marks, mark=*, mark size=2pt, orange, solid,
  error bars/.cd, y dir=both, y explicit,
  error bar style={line width=0.6pt, orange},
  error mark options={orange, line width=0.6pt},
] table[x=sigma, y=gc_Z16, y error=sd_Z16] {
sigma gc_Z16      sd_Z16
0.05  0.0312500   0.00465847
0.10  0.0322917   0.00389756
0.15  0.0322917   0.00208333
0.20  0.0364583   0.00465847
0.25  0.0437500   0.00255155
0.30  0.0489583   0.00255155
0.35  0.0510417   0.00389756
0.40  0.0531250   0.00607391
};
\legend{$Z=4$,$Z=6$,$Z=8$,$Z=16$}
\end{axis}
\end{tikzpicture}
\caption{Critical storage load $\gamma_c = P_c/(N d)$ versus angular
standard deviation $\sigma$ of the pattern-orientation distribution, for
different connectivities $Z$. Higher connectivity yields larger capacity
for all levels of orientational disorder, with $\gamma_c$ increasing
with $\sigma$. Error bars: standard deviation across $5$ replicas.}
\label{fig:orientation_effect}
\end{figure}

The critical coupling $\gamma_c$ depends on both orientational disorder
$\sigma$ and connectivity $Z$. As shown in
Fig.~\ref{fig:orientation_effect}, $\gamma_c$ increases with $\sigma$ for
each $Z$, with fluctuations due to finite sampling of $P_{\text{critical}}$.
Networks with higher connectivity consistently achieve greater memory
capacity: for $Z=4$, $\gamma_c$ ranges from $0.0104$ to $0.0156$, while
for $Z=16$ it increases from $0.0313$ to $0.0531$ over the same $\sigma$
range. Thus denser networks support higher baseline capacity and gain
more from orientational disorder.

The monotonic increase of $\gamma_c$ with $\sigma$ suggests that more
diverse orientations create a more robust memory landscape, reducing
interference between stored patterns via orientation-based
differentiation. This enhanced stability requires stronger synaptic
coupling, as reflected in higher $\gamma_c$ values. The variations of
$\gamma_c(\sigma)$ across different $Z$ indicate that connectivity
controls how orientational disorder affects pattern stability: sparse
networks require a minimum orientational diversity to overcome limited
connectivity, whereas dense networks exploit orientation diversity
effectively over the full parameter range.

The vectorial Hopfield model generates energy landscapes with multiple
local minima corresponding to stored patterns. We quantify pattern
stability via the energy difference between stored and random
configurations. For $N=25$ and $\gamma = 0.1$, stored patterns have mean
energies about $7$ units lower than random states, ensuring robust
retrieval against initialization noise. The energy function given by Eq. \eqref{energy}
naturally creates this hierarchy through the outer-product construction
of $\mathbf{W}$. The energy gap decreases as $\gamma$ approaches
$\gamma_c$, paralleling metastable hierarchies in glass-forming systems
\cite{Biroli,PhysRevB.101.014113,Charbonneau2014}.

The eigenvalue spectrum of $W_{ij}$ [Eq.~\eqref{weight}] characterizes
computational properties at different pattern densities (see
Fig.~\ref{spectral} in the End Matter). For low density ($\gamma = 0.05$),
the spectrum shows isolated pattern eigenvalues and a bulk near zero.
For intermediate density ($\gamma = 0.1$–$0.2$), a bulk develops with
pattern eigenvalues as outliers. For high density ($\gamma = 0.5$), the
spectrum approaches random-matrix behavior. The evolution of spectral
density with $\gamma$ reflects a crossover from a memory-dominated to a
noise-dominated regime; $\gamma_c$ corresponds to the merging of pattern
eigenvalues with the bulk and the loss of discriminability.

Convergence properties depend on initialization and pattern density. The
convergence time scales logarithmically with system size $N$ for
successful retrievals, while basin size decreases approximately
exponentially as $\gamma$ approaches $\gamma_c$, and basin boundaries
become more complex. The unit-vector constraint produces dynamics on the
product manifold $S^2 \times S^2 \times \cdots \times S^2$, preserving
pattern structure but adding geometric complexity. The update rule
\begin{equation}
\mathbf{s}_i^{(t+1)} = \frac{\mathbf{h}_i^{(t)}}{|\mathbf{h}_i^{(t)}|}, \quad \mathbf{h}_i^{(t)} = \sum_{j \neq i} \mathbf{W}_{ij} \mathbf{s}_j^{(t)}
\end{equation}
ensures convergence to fixed points while enforcing the spherical
constraint.

The block-structured matrix $\mathbf{W}$ encodes orientational
correlations through its $3 \times 3$ blocks $\mathbf{W}_{ij}$. We
characterize them via (i) the Frobenius norm $|\mathbf{W}_{ij}|_F$,
measuring interaction strength, and (ii) the anisotropy ratio
$\lambda_{\max}/\lambda_{\min}$ of $\mathbf{W}_{ij}^T \mathbf{W}_{ij}$,
measuring directional bias. For typical systems we find
$\langle|\mathbf{W}_{ij}|_F\rangle \approx 0.02$ and
$\langle\lambda_{\max}/\lambda_{\min}\rangle \approx 15$, indicating
strongly directional interactions due to the vectorial nature of stored
patterns. This correlation structure produces emergent organization in
$\mathbf{W}$, with stronger correlations between blocks associated with
patterns of similar orientation.

The vectorial Hopfield model introduced here extends associative memory
beyond binary states by assigning to each neuron a unit vector on $S^2$
and implementing block-structured couplings analogous to the Hessian of
disordered solids. This construction generates a rigid energy landscape
with well-separated minima associated with the stored patterns and
monotonic Lyapunov descent under the spherical constraint.

Our results show a clear performance gain over the classical binary
Hopfield network. For moderate connectivity, the critical storage ratio
$\gamma_c$ grows approximately linearly with $Z$, with dynamical and
spectral estimates in close agreement
[Fig.~\ref{fig:gamma_two_panels}]. The high-connectivity data
[Fig.~3] reveal a further enhancement: for large $Z$,
$\gamma_c$ systematically exceeds the extrapolated low-$Z$ linear trend,
indicating a crossover to an enhanced-capacity regime. This behaviour is
consistent with a connectivity-driven ``rigidification'' of the
associative landscape, whereby additional block couplings increase local
redundancy, strengthen effective fields, and reduce cross-talk.

Spectral analysis supports this interpretation: pattern modes remain
well separated from the bulk over an extended range of loads, and the
associated basins of attraction are larger and more stable, enabling
robust retrieval even from strongly corrupted initial conditions. These
features have no analogue in the standard Hopfield model at comparable
load, underscoring the advantage of vectorial degrees of freedom and
amorphous-solid-inspired architecture.

Promising extensions include nonlinear learning rules, sparse or hierarchical connectivities that retain the high-$Z$ advantage, and higher-dimensional state manifolds \cite{krotov}. Overall, the vectorial Hopfield framework provides a compact, high-capacity associative memory that markedly outperforms the classical binary model in the moderate- to high-connectivity regime.

\section*{Acknowledgments}
AZ gratefully acknowledges funding from the European Union through Horizon Europe ERC Grant number: 101043968
``Multimech''. During the preparation of this work we became aware of
similar observations in a vectorial spin-glass model \cite{Nicoletti}.

\section*{End Matter}

This appendix collects the numerical diagnostics supporting the results presented
in the main text. We report: (i) direct evidence of Lyapunov convergence;  
(ii) the spectral softening of the Hessian leading to the critical load;  
(iii) the evolution of the weight–matrix spectrum with increasing pattern
density; (iv) quantitative energy separation between memory and spurious minima.

\begin{figure}[H]
\centering
\begin{tikzpicture}
\begin{axis}[
  prl,
  xmin=0, xmax=23,
  ymin=-32, ymax=0.7,
  xlabel={Update step},
  ylabel={Energy},
  smooth,
]
\addplot+[prlblue, mark=*, mark size=1.8pt, smooth, tension=0.18]
table[row sep=\\]{
x  y \\
1   0.504331 \\
2  -25.2499  \\
3  -27.9052  \\
4  -28.8532  \\
5  -29.4204  \\
6  -29.8113  \\
7  -30.2077  \\
8  -30.4861  \\
9  -30.8211  \\
10  -30.9084  \\
11 -30.9846  \\
12 -30.9522  \\
13 -30.9752  \\
14 -30.9856  \\
15 -30.9904  \\
16 -30.9926  \\
17 -30.9936  \\
18 -30.9940  \\
19 -30.9942  \\
20 -30.9943  \\
21 -30.9944  \\
22 -30.9944  \\
23 -30.9944  \\
};
\end{axis}
\end{tikzpicture}
\caption{Monotonic decay of the Lyapunov energy $E(t)$ during synchronous
updates for $N=25$, $d=3$, $P=5$. The system converges rapidly to the
minimum associated with the retrieved memory.}
\label{fig:energy-convergence}
\end{figure}

\vspace{0.3cm}

The loss of stability of stored patterns is captured by the smallest eigenvalue
$\lambda_{\min}$ of the Hessian. The curve in Fig.~\ref{fig:lambda_min_vs_gamma}
shows the softening and zero-crossing of $\lambda_{\min}$ as the load $\gamma$
approaches the critical value.

\begin{figure}[H]
\centering
\begin{tikzpicture}
\begin{axis}[
  width=\columnwidth,
  height=0.75\columnwidth,
  xmin=0, xmax=0.55,
  ymin=-0.0015, ymax=0.005,
  grid=both,
  xlabel={$\gamma$},
  ylabel={$\lambda_{\min}$},
  ticklabel style={font=\small},
  label style={font=\small},
  extra y ticks={0},
  extra y tick style={grid=major, dashed, very thick},
]
\addplot+[
  only marks, mark=*, mark size=2.5pt, prlblue,
] table[x=gamma, y=lambda_min] {lambda_min_vs_gamma_Z63.dat};

\addplot[prlblue, thick, smooth]
table[x=gamma, y=lambda_min] {lambda_min_vs_gamma_Z63.dat};
\end{axis}
\end{tikzpicture}
\caption{Minimum eigenvalue of the Hessian for a fully connected
network ($Z=63$). The zero-crossing at $\gamma_c^{\text{spec}} \simeq 0.21$
marks the onset of instability of the memory attractors.}
\label{fig:lambda_min_vs_gamma}
\end{figure}

\vspace{0.4cm}

\begin{figure*}[p]
\centering
\begin{subfigure}{0.47\textwidth}
\includegraphics[width=\linewidth]{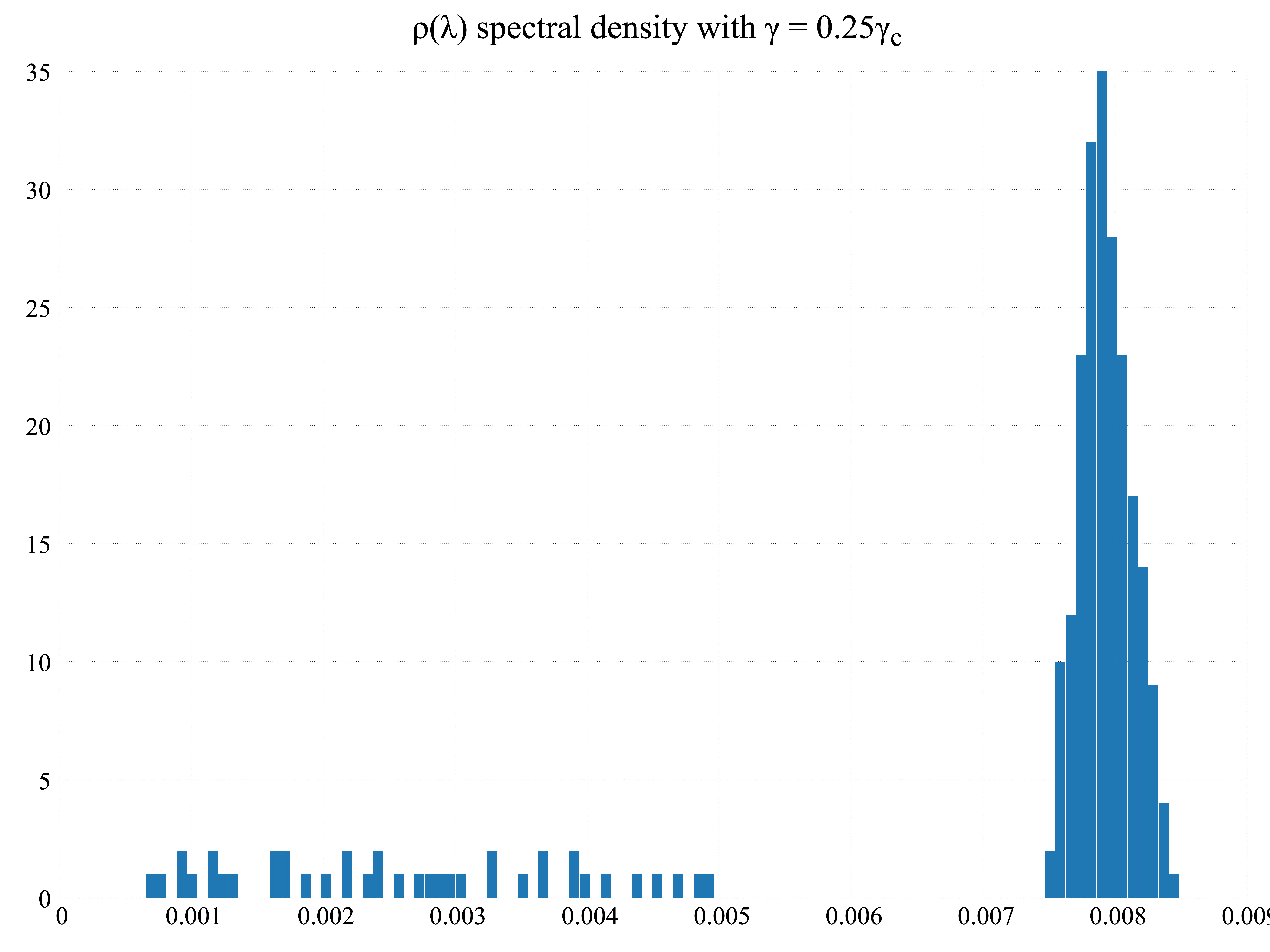}
\end{subfigure}
\hfill
\begin{subfigure}{0.47\textwidth}
\includegraphics[width=\linewidth]{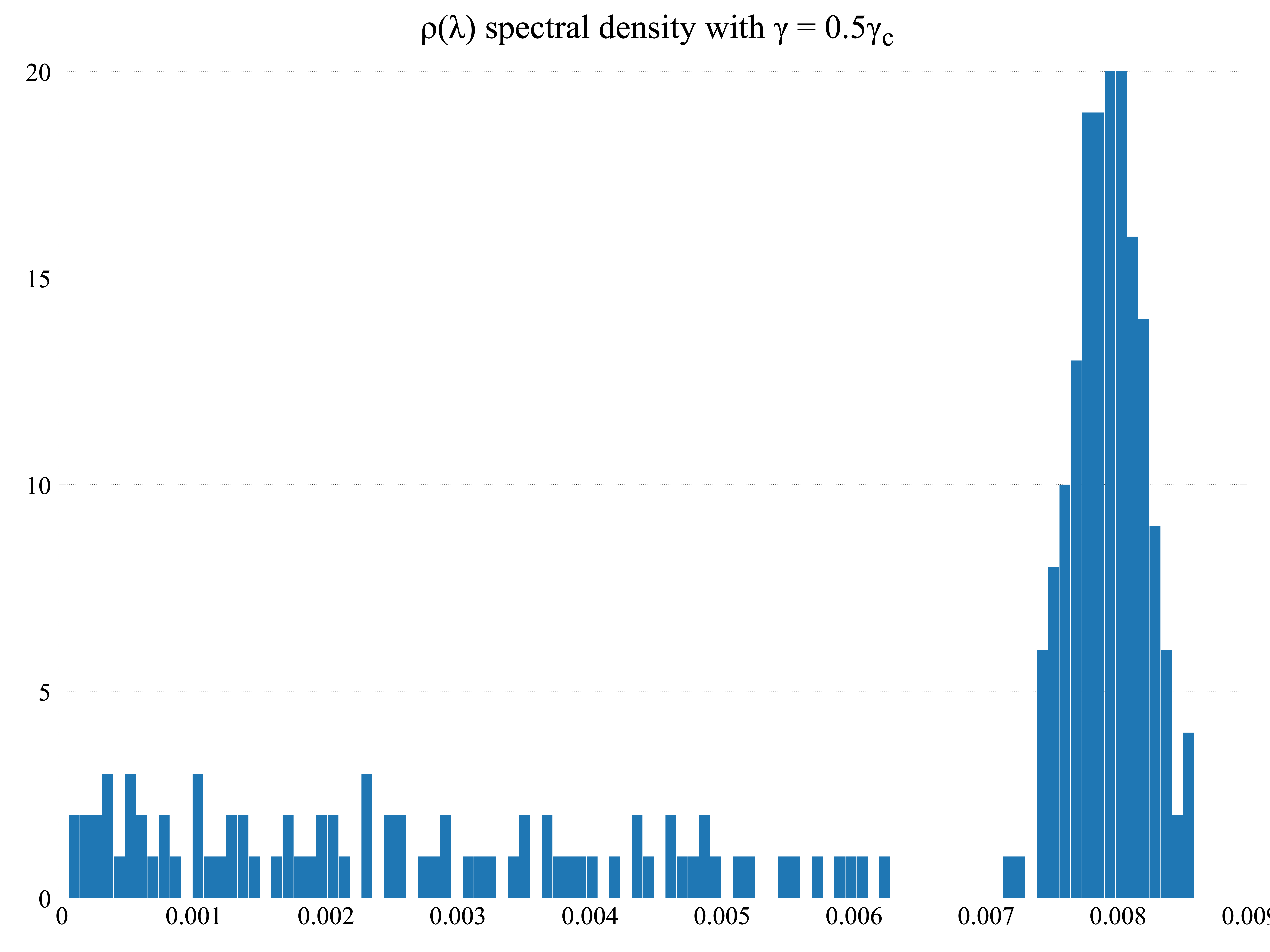}
\end{subfigure}

\vspace{0.75cm}

\begin{subfigure}{0.47\textwidth}
\includegraphics[width=\linewidth]{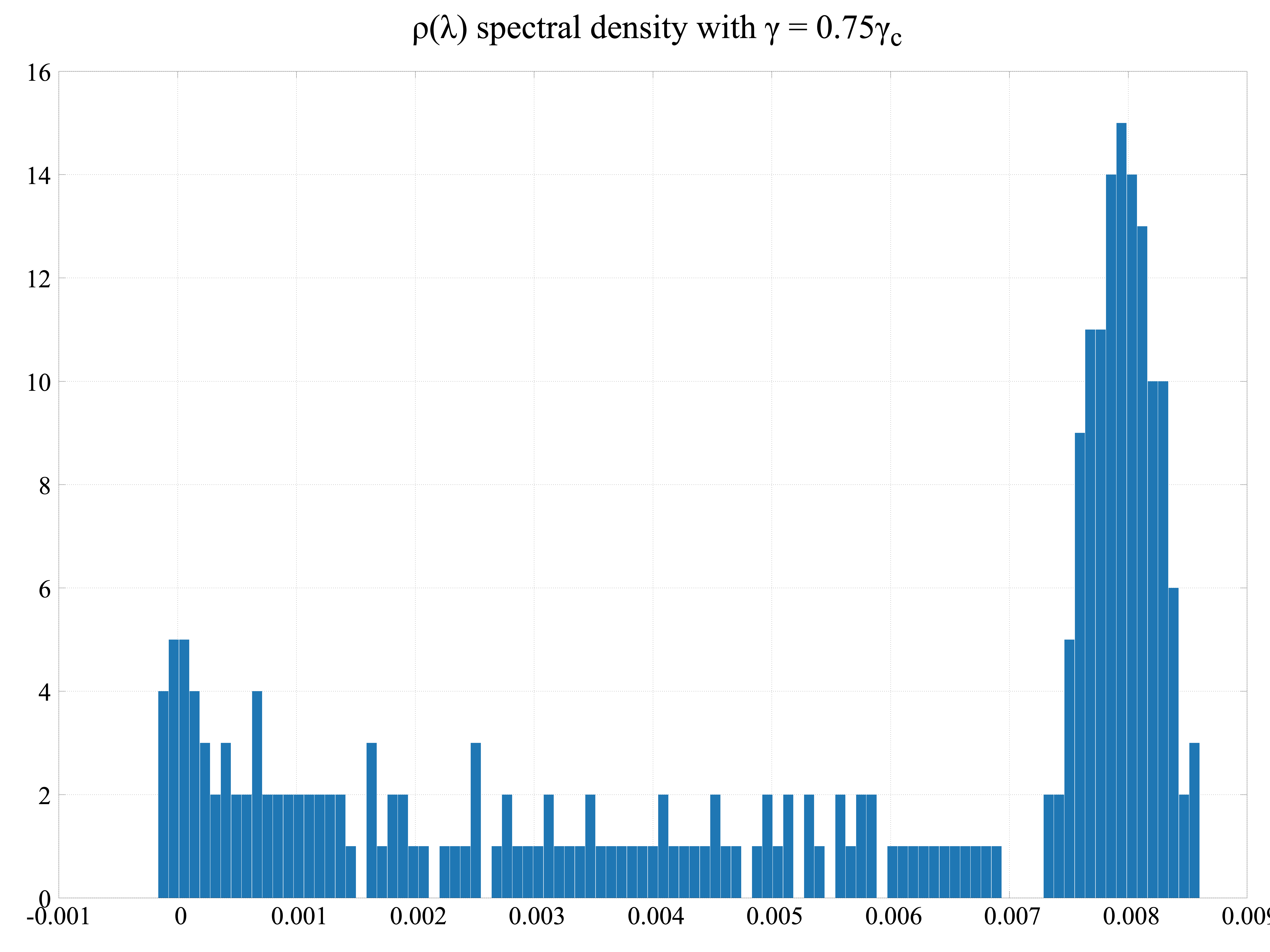}
\end{subfigure}
\hfill
\begin{subfigure}{0.47\textwidth}
\includegraphics[width=\linewidth]{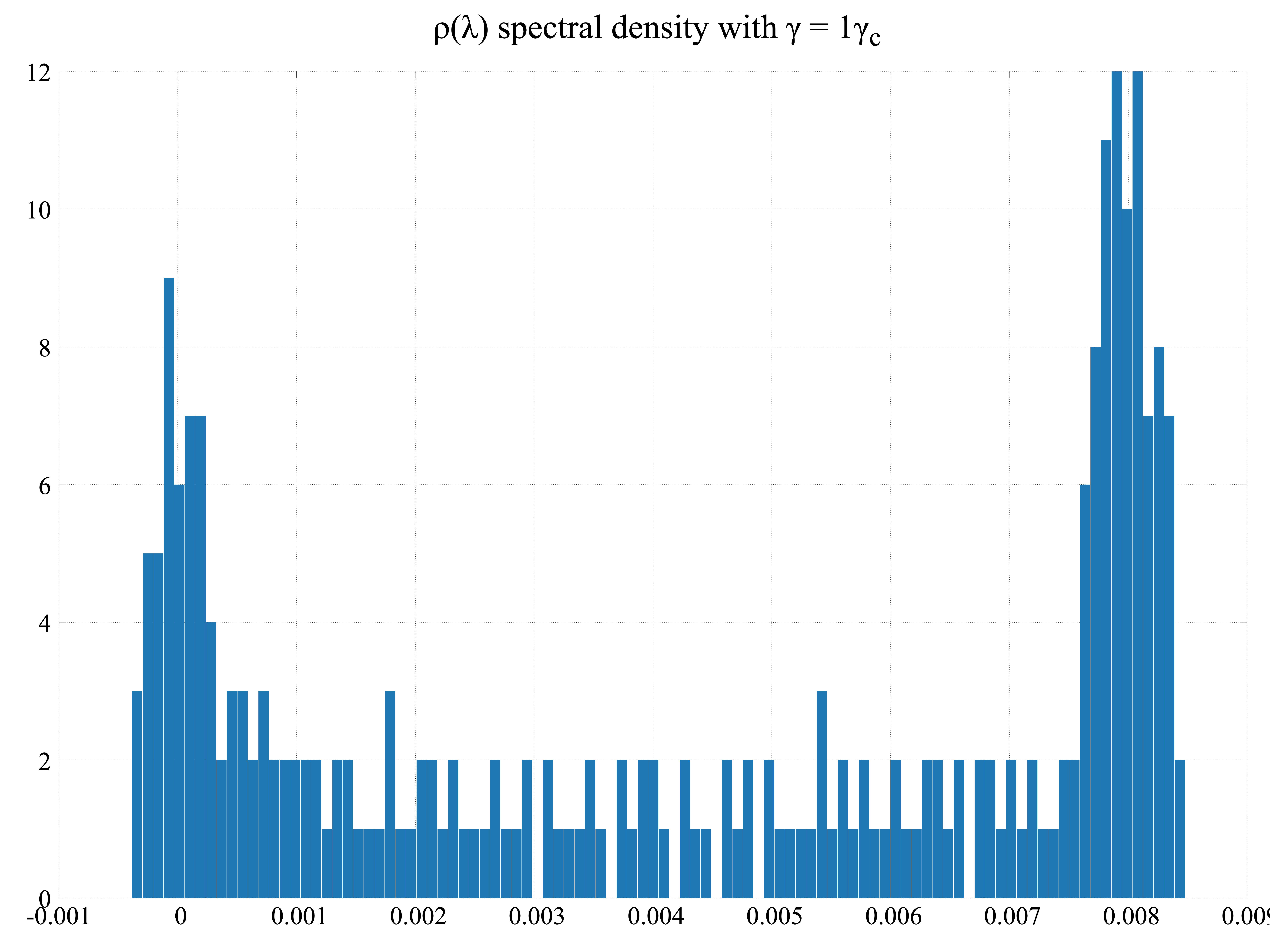}
\end{subfigure}

\caption{Evolution of the spectral density $\rho(\lambda)$ of the weight matrix
with increasing pattern load $\gamma$. Low loads (top left) exhibit isolated
pattern eigenvalues; intermediate loads develop a broadened bulk; high loads
(bottom right) show a continuous spectrum approaching random-matrix form.}
\label{spectral}
\end{figure*}

Figure~\ref{spectral} displays the spectral density of the weight matrix $W$ for
increasing pattern loads. As $\gamma$ grows, isolated outliers merge with the 
bulk and the spectrum broadens into a continuous band, consistent with the 
breakdown of retrieval capacity.

\vspace{0.4cm}

Finally, the table below summarises the energy separation between stored patterns
and spurious minima for three representative load levels. Energies are obtained
from 300 random initialisations for each value of $\gamma/\gamma_c$.

\begin{table}[H]
\centering
\begin{tabular}{c|c|c|c}
\hline
$\gamma/\gamma_c$ & $E_{\min}^{\text{mem}}$ & $E_{\min}^{\text{sp}}$ & Gap $\Delta E$ \\
\hline
0.1 & $-30.397$ & ---      & ---      \\
0.3 & $-27.681$ & $-27.199$ & $+0.482$ \\
0.7 & $-24.952$ & $-24.884$ & $+0.068$ \\
\hline
\end{tabular}
\caption{Energies of stored memories and deepest spurious minima.
Large gaps at low load reflect clean memory dominance; near capacity the gap
collapses, consistent with interference-driven instability.}
\label{tab:energy_gaps}
\end{table}

\bibliography{references}

\end{document}